\documentclass{nature}

\usepackage{xcolor}
\usepackage{soul}
\usepackage{graphicx}
\usepackage{float}

\newcommand{\DAQ}{WD\,J0551$+$4135}
\newcommand{\DAQlong}{WD\,J055134.612$+$413531.09}
\newcommand{\Msun}{\mbox{$\mathrm{M}_\odot$}}
\newcommand{\Rsun}{\mbox{$\mathrm{R}_\odot$}}
\newcommand{\Teff}{\mbox{$T_\mathrm{eff}$}}
\newcommand{\logZ}[1]{\mbox{$\log(\mathrm{#1}/\mathrm{H})$}}

\newcommand{\logmass}{\mbox{$-\log(1-m_r/M_\mathrm{wd})$}}

\newcommand{\Ion}[2]{\mbox{#1\,\textsc{#2}}}
\newcommand{\kms}{\mbox{km\,s$^{-1}$}}
\newcommand{\uHz}{\mbox{$\mu$Hz}}

\title{An ultra-massive white dwarf with a mixed hydrogen-carbon atmosphere as a likely merger remnant}

\author{
  M.~A.~Hollands$^{1}\star$,
  P.-E.~Tremblay$^{1}$,
  B.~T.~G\"ansicke$^{1}$,
  M.~E.~Camisassa$^{2,3}$,
  D.~Koester$^{4}$,
  A.~Aungwerojwit$^{5}$,
  P.~Chote$^{1}$,
  A.~H.~C\'orsico$^{2,3}$,
  V.~S.~Dhillon$^{6,7}$,
  N.~P.~Gentile-Fusillo$^{1,8}$,
  M.~J.~Hoskin$^{1}$,
  P.~Izquierdo$^{7,9}$,
  T.~R.~Marsh$^{1}$, \&
  D.~Steeghs$^{1}$
 }

\begin{document}

\maketitle

\begin{affiliations}
 \item Department of Physics, The University of Warwick, Coventry, CV4 7AL, UK
 \item Facultad de Ciencias Astr\'onomicas y Geof\'isicas, Universidad Nacional de La Plata,
       Paseo del Bosque s/n, 1900 La Plata, Argentina
 \item Instituto de Astrof\'isica de La Plata, UNLP-CONICET, Paseo del Bosque s/n, 1900 La Plata, Argentina
 \item Institut f\"ur Theoretische Physik und Astrophysik, University of Kiel,
       24098 Kiel, Germany
 \item Department of Physics, Faculty of Science, Naresuan University,
       Phitsanulok 65000, Thailand
 \item Department of Physics and Astronomy, University of Sheffield, Sheffield, S3 7RH, UK
 \item Instituto de Astrof\'isica de Canarias, 38205 La Laguna, Tenerife, Spain
 \item European Southern Observatory, Karl-Schwarzschild-Str 2, D-85748 Garching, Germany
 \item Departamento de Astrof\'isca, Universidad de La Laguna, 38206 La Laguna,
       Tenerife, Spain
\end{affiliations}

\begin{abstract}

White dwarfs are dense, cooling stellar embers consisting mostly of carbon and
oxygen\cite{pacz70}, or oxygen and neon (with a few percent carbon) at higher
initial stellar masses\cite{camisassa19}. These stellar cores are enveloped by
a shell of helium which in turn is usually surrounded by a layer of hydrogen,
generally prohibiting direct observation of the interior composition. However,
carbon is observed at the surface of a sizeable fraction of white
dwarfs\cite{hollandsetal18-2,kepler19}, sometimes with traces of oxygen, and it
is thought to be dredged-up from the core by a deep helium convection
zone\cite{koester82,pelletier86}. In these objects only traces of hydrogen are
found\cite{coutu19,koester19} as large masses of hydrogen are predicted to
inhibit hydrogen/helium convective mixing within the envelope\cite{rolland18}.
We report the identification of \DAQlong, an ultra-massive (1.14\,\Msun) white
dwarf with a unique hydrogen/carbon mixed atmosphere
($\mathrm{C}/\mathrm{H}=0.15$ in number ratio). Our analysis of the envelope
and interior indicates that the total hydrogen and helium mass fractions must
be several orders of magnitude lower than predictions of single star
evolution\cite{iben1983}: less than $10^{-9.5}$ and $10^{-7.0}$, respectively.
Due to the fast kinematics ($129\pm5$\,\kms\ relative to the local standard of
rest), large mass, and peculiar envelope composition, we argue that \DAQ\ is
consistent with formation from the merger of two white dwarfs in a tight binary
system\cite{toonen12,shen12,cheng19,Gvaramadze19}. \end{abstract}

\begin{figure}
    \centering
    \includegraphics[width=\columnwidth]{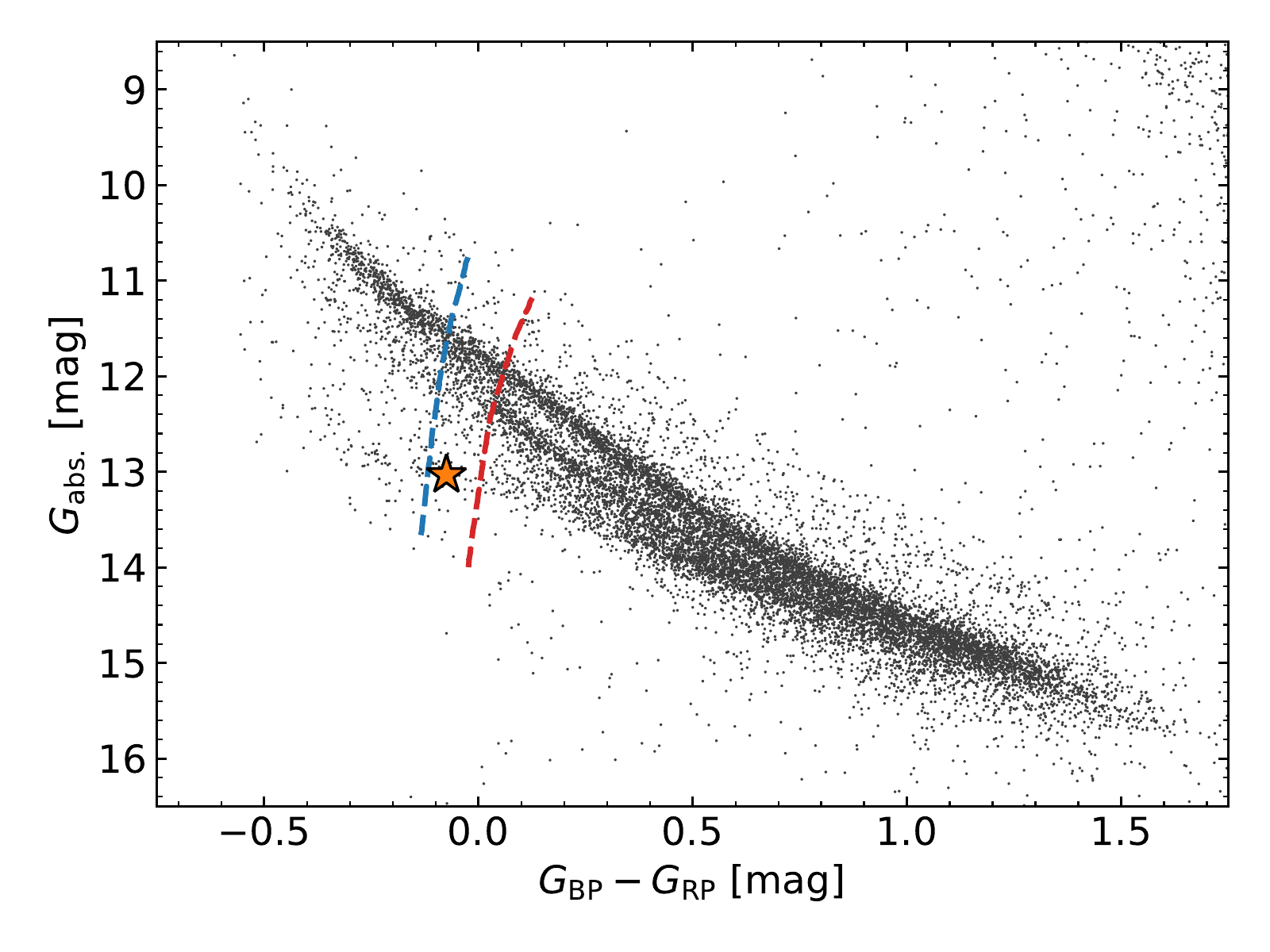}
    \caption{
    \label{fig:HRD}
    The white dwarf sequence (grey points) in the \textit{Gaia}
    Hertzsprung-Russell diagram. \DAQ\ -- indicated by the orange star -- is
    offset below (fainter than) the central locus of the white dwarf sequence,
    demonstrating its relatively small radius and thus high mass. The blue and
    red dashed curves bound the instability-strip for white dwarfs with
    hydrogen dominated atmospheres\cite{tremblayetal15-2}, defining the region
    in which they are unstable to pulsations. \DAQ\ is also located within the
    sequence of white dwarfs undergoing crystallisation\cite{tremblayetal19-1}.
    }
\end{figure}

\DAQ\ was identified as a candidate high-mass white
dwarf\cite{gentilefusilloetal19-1} from its location in the \textit{Gaia}
Hertzsprung-Russell diagram\cite{gaia1} (Fig.\,\ref{fig:HRD}). The
\textit{Gaia} parallax places \DAQ\ at $46.45\pm0.15$\,pc away from the Sun.
Compared to most other white dwarfs of the same $G_{BP}-G_{RP}$ colour, \DAQ\
is fainter in absolute magnitude, signifying a relatively small radius, and
thus a large mass given the mass-radius relation of white dwarfs whose
structure is dominated by electron degeneracy\cite{M-R}.

\newpage

\begin{figure}
    \centering
    \includegraphics[width=\columnwidth]{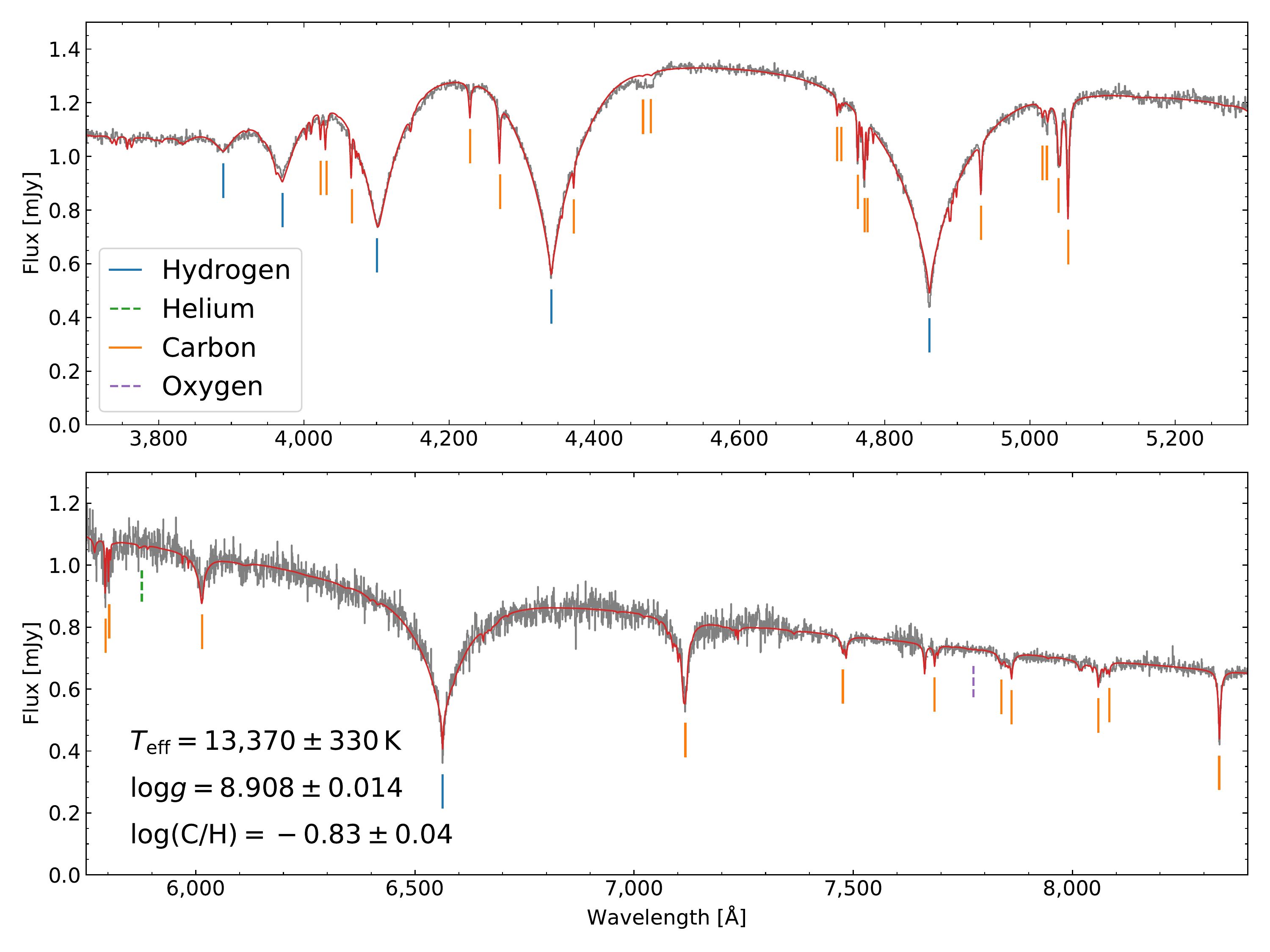}
    \caption{
    \label{fig:DAQspec}
    Combined optical spectrum from our observations of \DAQ\ (grey), with the
    best fitting model atmosphere in red. The hydrogen Balmer series is
    indicated in blue, whereas atomic carbon lines are labelled in orange. The
    expected locations of the strongest helium and oxygen features -- which are
    not detected -- are marked by the dashed green and purple lines
    respectively. Note that a few carbon lines (particularly towards the blue)
    are much weaker or absent in the data compared to the model, and so these
    were excluded in fitting the carbon abundance.
    }
\end{figure}

We acquired spectroscopy of \DAQ\ with the William Herschel Telescope (WHT) in
October 2018, February 2019, and September 2019. The coadded spectrum
(Fig.\,\ref{fig:DAQspec}) is qualitatively similar to a typical
hydrogen-atmosphere white dwarf, but with the addition of numerous absorption
lines from atomic carbon, formally making \DAQ\ the first white dwarf of its
spectral class\cite{sionetal83-1} (DAQ). Fitting (see Methods) the spectrum and
photometry (\textit{Gaia}/Pan-STARRS/\textit{SWIFT}), we found an effective
temperature (\Teff) of $13{,}370\pm330$\,K, radius ($R_\mathrm{wd}$) of
$(6.22\pm0.08)\times10^{-3}\,\Rsun$ (implying a mass ($M_\mathrm{wd}$) of
$1.140\pm0.008$\,\Msun, see Methods), and an atmospheric carbon to hydrogen
number ratio of $0.15\pm0.01$ ($\logZ{C} = -0.83\pm0.04$). The best fitting
model is shown in red in Fig.\,\ref{fig:DAQspec}. Helium is not detected in the
spectrum (Fig.\,\ref{fig:DAQspec}, green dashed) with a 99\,\% upper limit of
$\logZ{He} < -0.3$ (see Methods). This upper limit allows for a moderate helium
component, but confirms the hydrogen-dominated nature of the atmosphere. From
the absence of the 7,774\,\AA\ \Ion{O}{i} triplet (Fig.\,\ref{fig:DAQspec},
purple dashed), we determined an oxygen upper limit of $\logZ{O} < -4.5$ at
99\,\% confidence. Convection was identified from 0.02\,$\tau_R$ to the bottom
of the atmospheric model (1,000\,$\tau_R$), where $\tau_R$ is the Rosseland
mean optical depth.

\newpage

Previously known white dwarfs with helium/carbon atmospheres and similar
atmospheric parameters\cite{coutu19,koester19} are best explained by the
convective dredge-up of carbon into an outer helium
envelope\cite{koester82,pelletier86}. Assuming carbon dredge-up is also the
mechanism responsible for the nature of \DAQ, the helium fraction in the
stellar envelope must be exceptionally low -- dredge-up of carbon into the
surface hydrogen layer would necessarily also dredge-up all of the helium
situated between the carbon and hydrogen layers. The above
$\mathrm{C}/\mathrm{H}$ ratio is consistent with the highest $\mathrm{C/He}$
ratios observed in carbon-rich white dwarf atmospheres with comparable
\Teff\cite{koester19,coutu19}.

Given its mass, \DAQ\ is expected to harbour an ONe core, and so oxygen
dredge-up might also be expected, as is seen for some carbon rich white
dwarfs\cite{koester19,coutu19}. Since oxygen was not detected in the spectrum
of \DAQ, we considered the possibility that \DAQ\ harbours a massive CO core.

\begin{figure}
    \centering
    \includegraphics[width=\columnwidth]{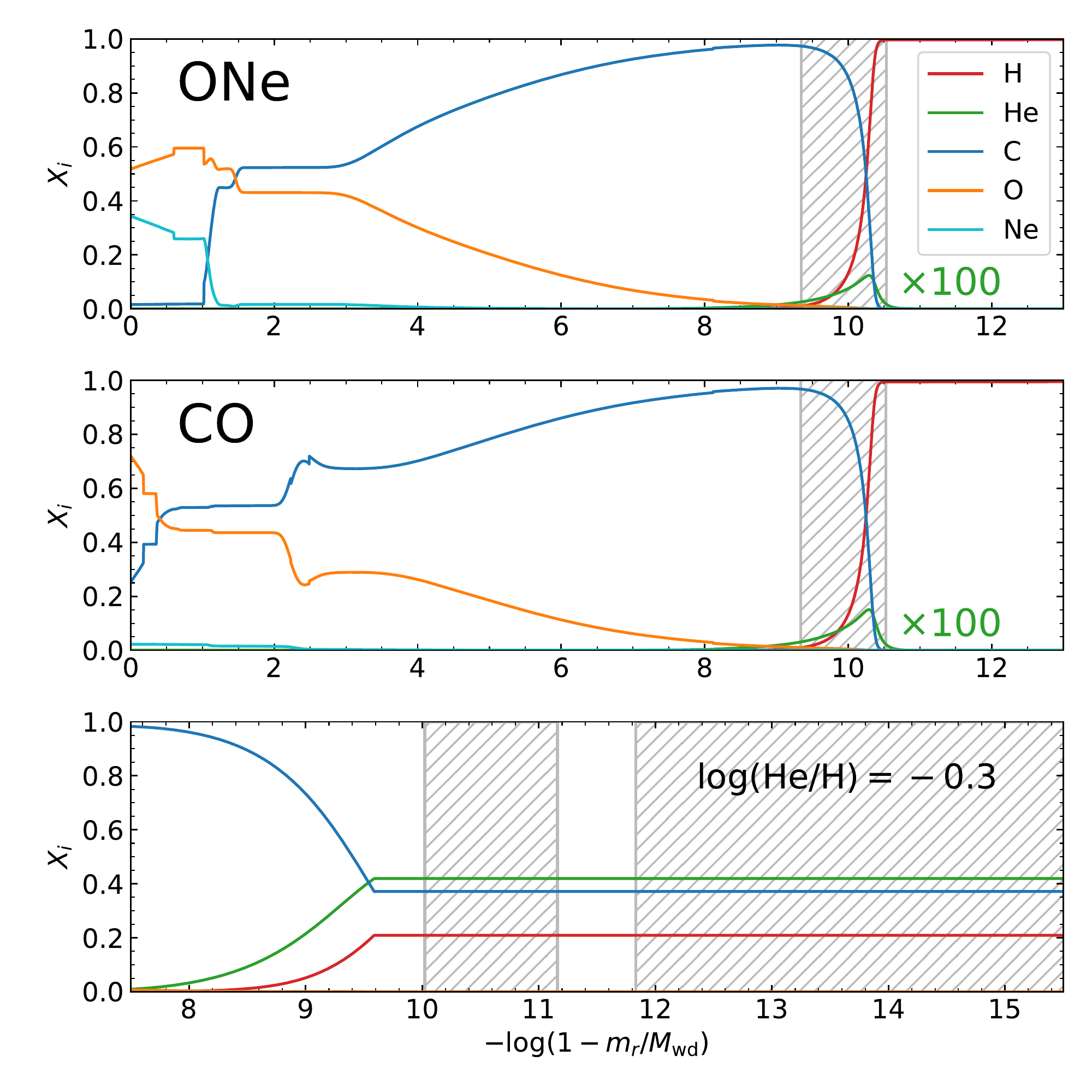}
    \caption{
    \label{fig:DAQinterior}
    Elemental mass fractions, $X_i$, against logarithmic mass-depth. Convection
    zones are hatched in grey. The top and middle panels show our interior
    models of \DAQ\ for ONe and CO cores respectively, with both hydrogen and
    helium masses fixed to $10^{-10}$\,\Msun\ and $\Teff=13{,}370$\,K. Helium
    curves have been multiplied by 100 for visibility. Chemical mixing is not
    included in these models. The bottom panel shows our 1D atmospheric model
    integrated downwards throughout the envelope, with the photospheric helium
    abundance set to the spectroscopic upper limit. Here, two convection zones
    are seen separated by 0.5\,dex in mass. Overshoot is assumed to extend
    chemical mixing one pressure scale-height either side of the 1D convection
    zones.
    }
\end{figure}

To test this hypothesis we calculated CO and ONe core interior models of \DAQ\
at the estimated mass and evolved them to the measured \Teff, using
\textsc{lpcode}\cite{camisassa19,althausetal12-1}. The envelope was initialised
with both hydrogen and helium masses of $10^{-10}\,M_\mathrm{wd}$ (chosen from
an initial estimate of the surface convection zone depth). The resulting models
shown in Fig.\,\ref{fig:DAQinterior} (top and middle panels) demonstrate
several key results. Firstly, despite the distinct core-structures, above
$\logmass=4$ (where $m_r$ is the mass enclosed within a sphere of radius $r$),
the models are almost indistinguishable. Therefore the degree of oxygen
dredge-up is not found to depend upon the core-composition, and so the
atmospheric non-detection of oxygen cannot be used to test the evolutionary
history.

Secondly, the helium layer, initially acting as a buffer between the hydrogen
and carbon, was found to have almost entirely diffused into the carbon layer --
multiplication of the helium-curves by $\times100$ is required for visibility
in Fig.\,\ref{fig:DAQinterior}. While the helium fraction peaks in the C/H
overlap region, most of the integrated mass is found between $\logmass = 2$ and
$7$, far below any convective region. This demonstrates that for low total
helium masses, diffusion succeeds in erasing the helium buffer, resulting in a
direct hydrogen/carbon interface, allowing for carbon dredge-up into the
hydrogen layer without dredging up a large amount of helium.

For a more accurate understanding of the upper layers, we integrated our best
fit model atmosphere downwards\cite{koester09-1} (Fig.\,\ref{fig:DAQinterior},
bottom), which explicitly includes chemical mixing in convective regions. The
displayed profile assumes our atmospheric upper limit of $\logZ{He}=-0.3$.
Below the surface convection zone, a second, deeper convectively unstable
region develops ($\logmass \in [9.6, 11.3]$). We obtained conservative upper
limits for the total hydrogen and helium mass fractions by integrating the
chemical profiles in the envelope model, taking into account convective
overshoot\cite{cunningham2019}, finding $10^{-9.5}$ and $10^{-7.0}$,
respectively (see Methods). While total hydrogen masses this low are observed
in a moderate fraction of white dwarfs, the canonical helium mass
fraction\cite{iben1983} is of the order $10^{-2}$, and so \DAQ\ is found to be
helium-deficient by many orders of magnitude.

We also investigated the kinematics of \DAQ. The proper motion and
(gravitational redshift corrected) radial velocity, imply a speed of
$118\pm5$\,\kms\ in the heliocentric frame, and $129\pm5$\,\kms\ in the local
standard of rest frame. We found this speed to be at the 99th percentile of the
3D velocity distribution of nearby white dwarfs with similar absolute
magnitudes (see Methods). Because stellar velocity dispersion increases with
system age, the fast kinematics of \DAQ\ may signify a system age much older
than implied simply from the white dwarf cooling: if \DAQ\ formed through
single star evolution, the implied progenitor mass would be
$6$--$7$\,\Msun\cite{cummingsetal18-1}, implying a main sequence lifetime of
$<10^8$\,yr, and thus a system age dominated by the estimated $1.3\pm0.1$\,Gyr
of white dwarf cooling. By comparison, formation from a double white dwarf
merger would allow \DAQ\ to be many Gyr older considering the longer main
sequence lifetimes of the lower mass progenitor stars, and the delay inferred
for merger products in the local white dwarf population\cite{cheng19}.

\begin{figure}
    \begin{centering}
    \includegraphics[width=0.95\textwidth]{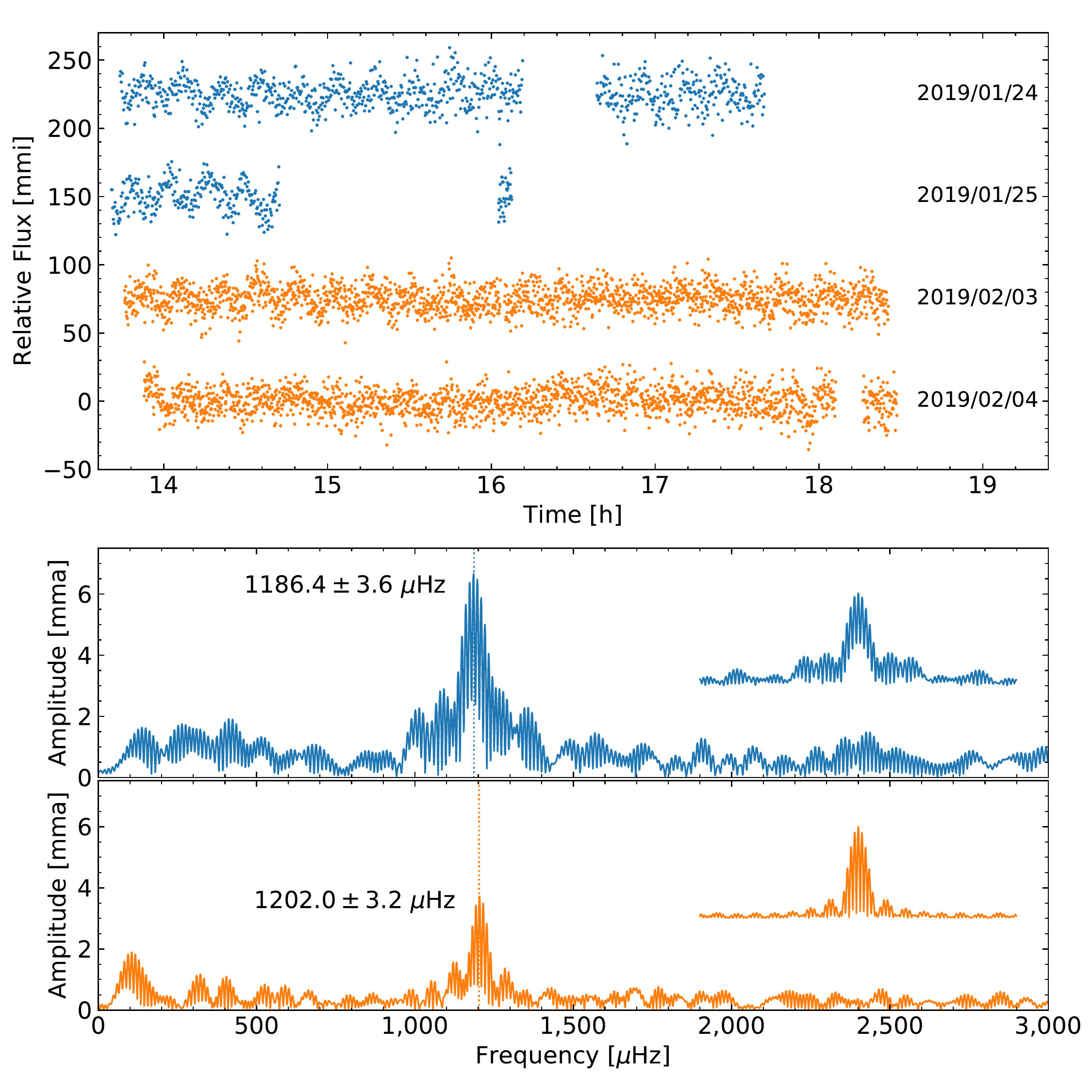}
    \end{centering}
    \caption{
    \label{fig:lcurve}
    Top: lightcurves for each night, offset by 75\,milli-modulation intensity
    (mmi) from one another. Time refers to the elapsed time since the start of
    the barycentric Julian date (BJD) of each night. Bottom: Fourier Transforms
    for each pair of nights, with colours corresponding to the above
    lightcurves. A single peak is evident in each, which is found to change in
    both frequency and amplitude. The spectral window is shown in the top-right
    of each panel.
    }
\end{figure}

The mass and temperature of \DAQ\ locates it in the parameter-space where white
dwarfs with hydrogen dominated atmospheres are unstable to
pulsations\cite{tremblayetal15-2} (bounded by the blue and red dashed lines in
Fig.\,\ref{fig:HRD}), motivating a search for pulsations in \DAQ. Observations
obtained with the Liverpool Telescope indicated potential variability at a
$\approx850$\,s period $\approx1\,\%$ amplitude. Additional lightcurves
(Fig.\,\ref{fig:lcurve}, top) obtained with the Thai National Telescope (TNT)
-- two nights in January and two in February 2019 -- unambiguously confirmed
the variability as a single pulsation mode. The 840\,s period is found to be
typical for hydrogen-atmosphere white dwarfs with these stellar
parameters\cite{curdetal17-1}. While a single peak in the Fourier transform
(Fig.\,\ref{fig:lcurve}, bottom) may arise from rotation, the oscillation
frequency exhibits a $15.6\pm4.8$\,\uHz\ shift between the January and February
observations (see Methods), which is possible for pulsations, but not for
rotation. The oscillation amplitude also varies by a factor of two between the
observations, which could be attributed to unresolved rotational splitting.

Given the many unusual properties of \DAQ, this object is challenging to
explain via single star evolution. Instead we argue that its unique
hydrogen/carbon atmosphere, inferred low total hydrogen and helium content,
fast kinematics consistent with an old age, and finally the mass approximately
twice the canonical mass are consistent with formation from a double white
dwarf merger. Following such a merger, \DAQ\ might have had a hot carbon
dominated atmosphere\cite{dufouretal07-1} with hydrogen later diffusing upward
as the white dwarf cooled. 

If \DAQ\ formed from a merger, it can be expected to be rapidly rotating. While
rotational broadening of spectral lines was not detected ($v\sin i < 50$\,\kms,
where $v$ is the equatorial rotation speed, and $i$ is the inclination angle),
the tiny radius of \DAQ\ only allows the period to be constrained to $>9$\,min
(for edge on rotation). For example EUVE\,J0317$-$855 is also thought to be a
merger remnant, but its measured rotation period of
12\,minutes\cite{ferrarioetal97-1} would be undetectable for \DAQ. 

A possibility to distinguish between single star evolution and the merger
hypothesis comes from the core composition, where ONe and CO cores are expected
respectively. While interior modelling of \DAQ\ proved insufficient to
elucidate the core composition, a preliminary investigation (using the
\textsc{lp-pul} pulsation code\cite{corsico06}) into the pulsation properties
of our CO/ONe interior models suggests a dependence on core composition, and
has similarly been shown in other recent work\cite{degeronimoetal19-1}. Indeed
asteroseismology is the most successful methodology for investigating the
interior physics of white dwarfs, including core
composition\cite{corsicoetal19-1}. Therefore the future detection of additional
pulsation frequencies may allow the core composition to be established via
detailed asteroseismology, and the formation channel to be resolved.

\vspace{-3mm}
The unique atmosphere of \DAQ\ has been shown to be explainable if both
hydrogen and helium masses are orders-of-magnitude lower than usual, where the
helium layer has mostly diffused downwards, allowing carbon dredge-up into the
hydrogen layer. These unusually low hydrogen/helium masses could be the result
of a merger of a former binary system -- a hypothesis supported by the fast
kinematics and ultra-massive nature of \DAQ. However, the nature of the helium
depleting process remains an outstanding question of this work, and so
motivates exploration using detailed white dwarf merger simulations including
helium and hydrogen burning\cite{shen12,Tanikawa15}.


\begin{methods}

\subsection{Spectroscopic observations and reductions}

On October 15th 2018, we observed \DAQ\ as part of our International Time
Programme (ITP), using the William Herschel Telescope (WHT) at the Roque de los
Muchachos Observatory. We made our observations with the
Intermediate-dispersion Spectrograph and Imaging System (ISIS), which allows
simultaneous observations using blue and red optimised CCDs via a dichroic
beam-splitter. For the blue arm we used the R600B grating at a central
wavelength of 4,540\,\AA\, and R600R at 6,560\,\AA\ in the red arm, permitting
a spectral resolution of $\approx 1$\,\AA.

We took single 600\,s exposures in both arms commencing at 05:58 UTC. As the
final target of the night as well as the observing run, we did not have time to
obtain additional spectra either in this setup or at other central wavelengths.
Bias and flat field frames were taken in the afternoon before the start of
observing. Standard star (G\,191$-$B2B) and arc calibration data were taken
following observation of \DAQ.

Reduction of the data was performed using packages within the
\textsc{starlink}\cite{currieetal14-1} distribution of software. Standard
techniques were used, i.e. bias subtraction, flat field correction, sky
subtraction, and optimal extraction (\textsc{figaro}, \textsc{kappa}, and
\textsc{pamela} packages). Wavelength and flux calibration of the 1D spectra
were performed using \textsc{molly}\cite{marsh19-1}

Additional spectroscopic observations were acquired on February 9th 2019 and
September 7th 2019, again using ISIS on the WHT, once again using the R600B and
R600B gratings. For the February observations the central wavelengths were
instead set to 3,930\,\AA\ and 8,200\,\AA. Two 900\,s exposures were taken with
each CCD with arcs acquired in between. Compared to the October observations,
the February spectra have a more accurate wavelength calibration, suitable for
precise measurement of the redshift of \DAQ. G\,191$-$B2B was again used as a
flux standard. The September observations were acquired in service mode with
central wavelengths of 4,540\,\AA\ and 8,200\,\AA. Reduction of these data
followed the same procedure as described above, though for the September
observations a flux standard was not provided, and so the data were flux
calibrated against the best fitting model to the previous data. A summary of
all spectroscopic observations is given in Table~E\ref{tab:obslog}.

\subsection{Spectrophotometric fitting}

To determine the stellar and atmospheric parameters (summarised in
Table\,E\ref{tab:spectrophotofit1}), we made use of the multiple data available
to us, i.e. the spectroscopy, photometry, and the \emph{Gaia} parallax, fitting
these with the Koester white dwarf 1D model atmospheres\cite{koester10-1} and a
mass-radius relation. The fitting was split into two parts. Firstly, to fit the
\Teff\ and stellar radius, we used only the photometry and parallax, with fixed
abundances. For the second part we fitted the abundances using only the
spectra, but with \Teff\ and the radius fixed. The whole procedure was then
performed iteratively until all parameters converged to within a small fraction
of the quoted uncertainties. In both cases, the fits were achieved via $\chi^2$
minimisation. These photometric and spectroscopic fits are detailed below.

For the photometric fit, the choice of photometry and treatment of their
uncertainties are important. For the optical we used
\emph{Gaia}\cite{gaia1,gaia2} and Pan-STARRS\cite{Pan-STARSS1} which have been
shown to have consistent calibration in the case of warm white
dwarfs\cite{gentilefusilloetal19-1}. For the near-ultraviolet (NUV) we
initially made use of \emph{Galex} photometry\cite{galex}. While we were able
to achieve an adequate fit to these photometry, we found the resulting model
($\Teff = 12{,}400$\,K) proved inconsistent with the Balmer lines in our
optical spectroscopy. Suspecting poor absolute calibration of the \emph{Galex}
fluxes\cite{walletal19-1}, we acquired NUV \emph{SWIFT} photometry in December
2018. These observations were performed with the U, UVW1, and UVW2 filters
using the UVOT (Ultraviolet/Optical Telescope) instrument. To derive calibrated
AB magnitudes, we used the latest zeropoints from  the High-Energy Astrophysics
Science Archive Research Center (HEASARC) for all three filters. The flux level
was indeed found to be higher compared to \emph{Galex}
(Fig.\,E\ref{fig:photfit}). Refitting using the \emph{SWIFT} photometry in the
NUV instead, we were able to achieve a fit consistent with both photometry and
spectroscopy. All photometry considered are listed in
Table\,E\ref{tab:phot_ast_table}. To correctly fold the parallax and its
uncertainty into our fit, we included parallax as a free parameter, the
measured value and its error were then used as a Gaussian prior. Given the
proximity of \DAQ\ ($46.45\pm0.15$\,pc), the effects of interstellar reddening
could be neglected throughout, and 3D effects from
convection\cite{tremblayetal13-1} can also be neglected when fitting optical
photometry. Throughout a mixing length parameter of ML2/$\alpha$ = 1.25 was
used, though we note that little difference was seen for a value of 0.8.

\begin{sfigure}
    \centering
    \includegraphics[width=\columnwidth]{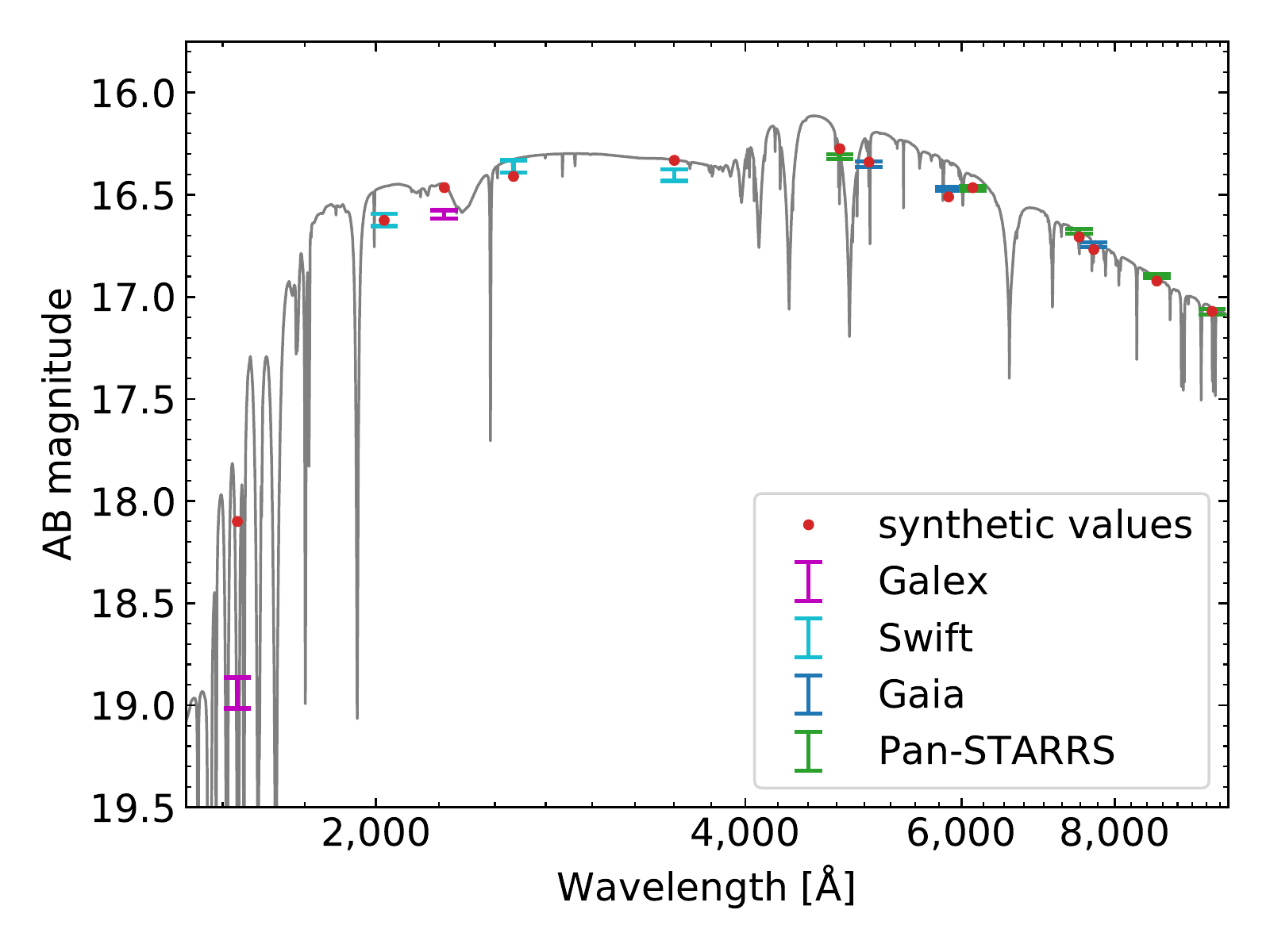}
    \caption{
    \label{fig:photfit}
    Our model spectrum (grey) was fitted to \emph{Gaia}, Pan-STARRS and
    \emph{Swift} photometry to determine the \Teff\ and stellar radius. Fitting
    the \emph{Galex} photometry instead of \emph{Swift}, a cooler \Teff\ of
    $12{,}400$\,K was found to be inconsistent with the optical spectrum. The
    \emph{Galex} magnitudes are therefore shown only to demonstrate
    disagreement with the superior absolute calibration of \emph{Swift}
    photometry. Error-bars represent $1\sigma$ uncertainties.
    }
\end{sfigure}
\newpage

For the spectroscopic fit, because the surface gravity is the input parameter
to the spectral synthesis, the stellar radius (determined from the photometric
fit) must be converted to a $\log g$ via a mass-radius relation. For this we
made use of CO models with thin hydrogen envelopes\cite{fontaineetal01-1},
linearly interpolating $\log g$ as a function of $\log R_\mathrm{wd}$ and $\log
\Teff$. This also leads to a mass estimate of $1.145\pm0.007$\,\Msun. However,
since we also consider an ONe-core as a possibility, we considered the effect
this would have on the mass. From our interior models (below), we determined an
ONe-core white dwarf would have a 0.85\,\% lower mass at our measured radius,
i.e. $1.136\pm0.007$\,\Msun. As compromise between these two possibilities, we
adopt $1.140\pm0.008$\,\Msun\ as the mass of \DAQ, and a surface gravity of
$\log g = 8.908\pm0.014$.

For measuring the carbon abundance not all lines were suitable. Bluer than
4,300\,\AA, the \Ion{C}{i} lines are overly strong in the model, whereas the
doublet near 4,570\,\AA\ is too weak (Fig.\,\ref{fig:DAQspec}). These lines are
reported with NIST (National Institute of Standards and Technology) grade E
(oscillator strength accuracy level $>50$\,\%). We therefore only considered
lines with grade B (7--10\,\% accuracy) or above, although only the 8,335\,\AA\
line was found to be higher accuracy with grade B+ (3--7\,\% accuracy) The
final model is shown with the data in Fig.\,\ref{fig:DAQspec}.

Owing to the sharp Balmer line cores and carbon lines, we estimate an upper
limit of 50\,kG on magnetic field strength for \DAQ.

\subsection{Upper limits}

In spectroscopic analyses upper limits are typically expressed as a confidence
level in percentage form, but usually without qualification of how such a
percentage was derived or should be interpreted. The helium/oxygen upper limits
we present are crucial to our interpretation of \DAQ\ and so here we make clear
our approach to their derivation.

For each of helium and oxygen, we used only their strongest predicted features
to assess upper limits -- the 5,877\,\AA\ \Ion{He}{i} line and the 7,774\,\AA\
\Ion{O}{i} triplet (which would appear unresolved if present in our data). For
He we created a grid of models from $\logZ{He} = -3.0$ to $+0.2$ in steps of
0.1\,dex with $\Teff$, $\log g$, and $\logZ{C}$ fixed to the values established
from our fits. For O we instead used a grid from $\logZ{O} = -7.0$ to $-3.6$
also in steps of 0.1\,dex. For each model we normalised them to the data by
fitting a first order polynomial to the surrounding continuum. We then
determined the $\chi^2$ between the models and data at each grid point.

The $\chi^2$ values can be converted to likelihoods via
\begin{equation}
    L \propto \exp(-\chi^2/2).
\end{equation}
Because the models barely change for increasingly negative log-abundances, the
likelihood converges to a constant value. This makes the likelihood problematic
for estimating an upper limit as this represents an improper distribution, one
that can't be normalised, and thus one where percentiles cannot be established.
This issue can be rectified through a careful choice of uninformative prior.

First we consider the abundances on the absolute fractional scale. For a
two-component atmosphere with abundances of hydrogen and a second element $Z$
(where $\mathrm{H} + Z = 1$), the likelihood becomes a proper distribution when
expressed as a function of absolute fractions, however, we still require an
explicit choice of prior. Given that we are attempting to constrain an unknown
fraction, a beta-distribution is the appropriate choice. We used the Jeffrey's
prior, setting the $\alpha$ and $\beta$ parameters of the beta distribution to
$1/2$, i.e.
\begin{equation}
    P(Z) = Z^{-1/2} (1-Z)^{-1/2} = Z^{-1/2} \mathrm{H}^{-1/2}.
    \label{eq:beta}
\end{equation}
For an $N$-component atmosphere, this can be extended to the corresponding
Dirichlet-distribution,
\begin{equation}
    P(\mathbf{Z}) = \prod_i^N Z_i^{-1/2},
    \label{eq:dirichlet}
\end{equation}
however with the C abundance fixed, equation~(\ref{eq:beta}) is sufficient.

With our model grids spaced equidistantly in log-abundance, numerical
integration is simpler on this scale rather than in terms of absolute
fractions. Transforming equation~(\ref{eq:beta}) we find
\begin{equation}
    P(\log(Z/\mathrm{H})) \propto \Big\{1+\cosh\big[\ln(10)\,\log(Z/\mathrm{H})\big]\Big\}^{-1/2},
    \label{eq:logbeta}
\end{equation}
as the corresponding prior for log-abundances. Note that in the limit
$Z \ll \mathrm{H}$
\begin{equation}
    P(\log(Z/\mathrm{H})) \propto 10^{\log(Z/\mathrm{H})/2},
    \label{eq:logbetaapprox}
\end{equation}
provides a simple approximation to (\ref{eq:logbeta}). Multiplying the
likelihood by the prior distribution given by (\ref{eq:logbeta}) yields a
proper distribution that initially rises with increasing abundance, before
rapidly dropping to near-zero posterior probability once the models start to
visibly disagree with the data. Finding a given percentile of the distribution
is a simple case of numerical integration, to obtain the cumulative
distribution and reading off the corresponding percentile. While we quoted the
99th percentiles, we note that for both He and O, the 95th percentiles were
found to be 0.2\,dex below the 99th percentile. The model spectra evaluated at
these upper limits are shown against the data in Fig.~E\ref{fig:uplims}.

\begin{sfigure}
    \centering
    \includegraphics[width=\columnwidth]{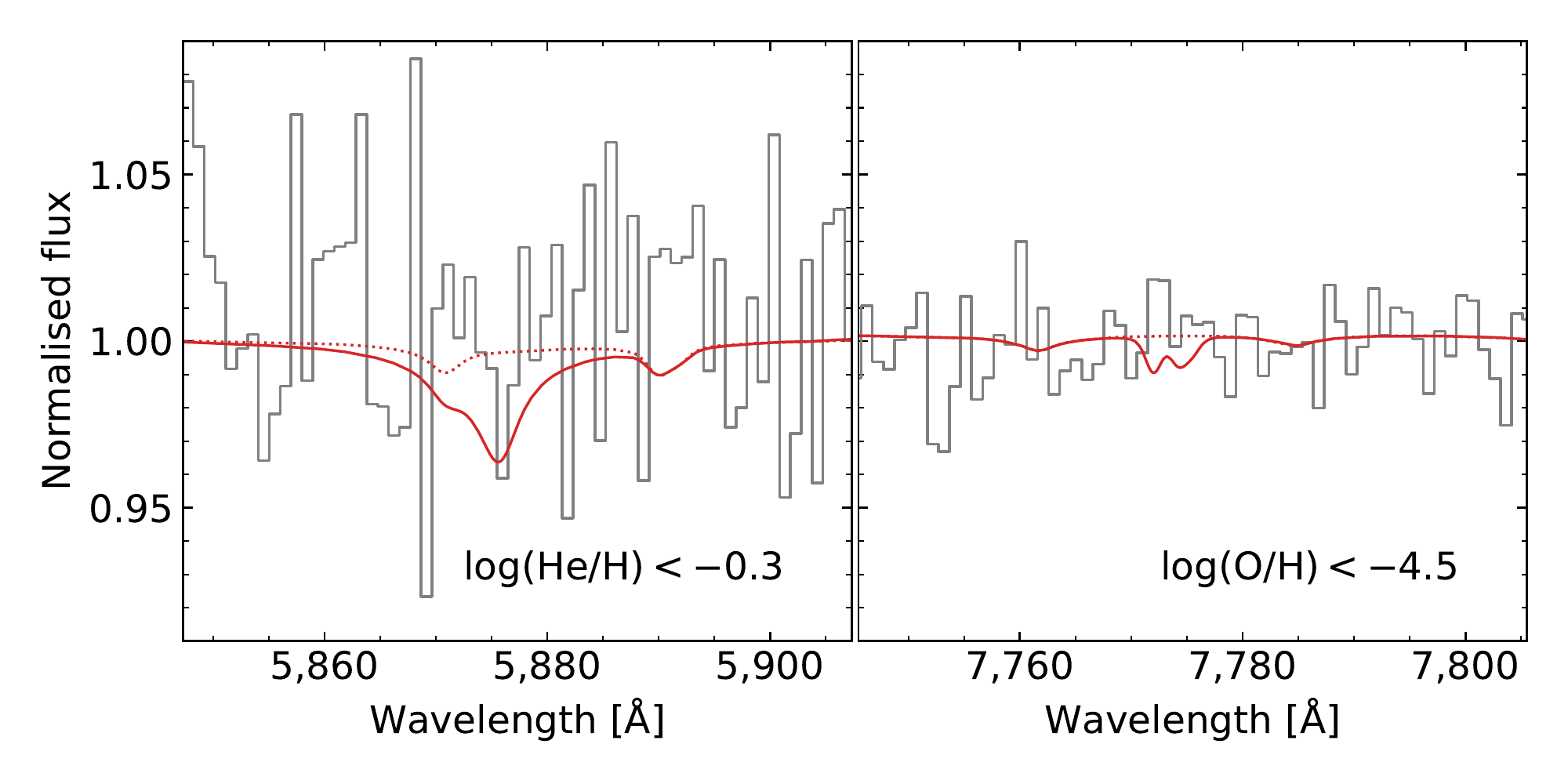}
    \caption{
    \label{fig:uplims}
    Upper limits for He and O abundances. The solid red models correspond to
    the estimated 99th percentile upper limits, whereas the dotted curves
    indicate models with their respective elements at zero abundance.
    }
\end{sfigure}

\subsection{Interior structure}

To analyse the unusual surface abundances of this star, we have performed
evolutionary calculations for both ONe-core and CO-core white dwarf models,
with low H and low He content, using the \textsc{lpcode} stellar evolutionary
code. While a merger origin may initially produce an out-of-equilibrium
structure, these evolutionary calculations are justified given that the
timescale for carbon diffusion out of the convection zone is of the order of
hundreds of years\cite{cunningham2019} for the stellar parameters of \DAQ. It
is therefore almost impossible that the presence of carbon is due to the
relaxation of prior evolution instead of dredge-up. The results are shown on
the top two panels of Fig.\,\ref{fig:DAQinterior} for the evolution stage that
corresponds to the mass and \Teff\ of \DAQ. The initial ONe chemical profile is
the result of the full previous calculation of the progenitor
evolution\cite{siess07-1,siess10-1}, whereas the initial CO chemical profile
was artificially generated. For the envelope we initialised both the hydrogen
and helium masses to $10^{-10}\,M_\mathrm{wd}$. These theoretical white dwarf
models consider changes in the chemical abundances due to chemical diffusion
and gravitational settling\cite{paquette1,paquette2,pelletier86}, and a
detailed treatment of the chemical redistribution resulting of crystallisation,
due to coulomb interactions\cite{camisassa19}. As a result of gravitational
settling, heavy elements are rapidly depleted from the outer layers, sinking
into the stellar interior. Simultaneously, chemical diffusion acts in the
opposite direction, smoothing the chemical profiles. 

Despite the distinct structures within the cores, we have found that the models
are almost indistinguishable in the envelope ($\logmass > 4$). However, the
effect of the low light element content produces notable differences in the
stellar envelope compared to previously published models with thick He- and
H-layers\cite{camisassa19}. Due to the low helium content of this star,
chemical diffusion manages to erase the helium buffer, leading to a
hydrogen-carbon interface, regardless of the core chemical composition.
Moreover, as helium and carbon have the same mass-to-proton ratio, electron
degeneracy does not prevent helium chemical diffusion into the interior, as it
does happen with hydrogen\cite{iben+mcdonald85-1}. Thus the hydrogen-carbon
interface can allow convection to dredge-up carbon, without revealing
signatures of helium in the spectrum. We note that these \textsc{lpcode} models
include convective energy transfer, with the inferred convective layers shown
in Fig.\,\ref{fig:DAQinterior}, but not the possibility of convective mixing.
Therefore to study the envelope in more detail, we rely on our best fit model
atmosphere that we integrate downwards using the same microphysics as that used
in the model atmosphere code\cite{koester09-1,koester10-1}. In particular, we
rely on a mixed H/He/C equation-of-state allowing for the different chemical
constituents to influence the size of the convection zone(s). As with the
atmospheric models, 1.25 was used for the mixing length parameter.

The integrated envelope model assumes complete convective mixing within the
unstable regions but neglects convective overshoot\cite{overshoot1,overshoot2}.
In radiative layers, microscopic diffusion is taken into account as in the
\textsc{lpcode}, allowing for a chemical gradient in which the fraction of
light elements decrease with increasing depth. We rely on two assumptions on
the surface composition, either mixed H/C or mixed H/He/C using our upper limit
on the helium abundance. Only the latter case is shown on the bottom of
Fig.\,\ref{fig:DAQinterior}, as it also leads to an upper limit on the total
mass of hydrogen and helium. The main difference between the He-free case and
our $\logZ{He} = -0.3$ model is that the surface convection zone is
significantly smaller in the former case, with its base located at $\logmass =
13.5$. However in both cases, a second, deeper unstable region develops
($\logmass \in [9.6, 11.3]$). This is in stark contrast with pure-hydrogen
structures which never develop a second, deeper convection
zone\cite{tremblayetal15-3}.

To obtain upper limits for the hydrogen and helium total mass fractions, we
must also consider the effect of convective overshoot. Thus we can not directly
rely on the 1D abundance profiles presented in the bottom of
Fig.\,\ref{fig:DAQinterior}. In the limiting case of $\logZ{He} = -0.3$ shown
in Fig.\,\ref{fig:DAQinterior}, the separation between the two convection zones
is small enough that overshoot is expected to be able to completely mix
them\cite{cunningham2019}. While 3D overshoot simulations for white dwarfs have
so far only been performed for pure-H and mixed H/He
atmospheres\cite{kupka18,cunningham2019,cukanovaite2019}, the presence of
carbon is not expected to change the strength of overshoot for a given
convection zone mass. Furthermore, convective overshoot is also expected to mix
material below the second convection zone. Here we apply a conservative one
pressure scale height value (0.43\,dex in mass). Therefore, we obtain our upper
limit by first integrating our envelope model assuming the atmospheric
composition down to 0.43\,dex below the second convection zone ($\logmass =
9.6$). Below this point, the diffusion tails in the interior models were
rescaled to match the value at $\logmass = 9.6$. This results in 99.8\,\% of
the helium mass being localised to the diffusion tail below the convectively
mixed region. This ultimately yields $10^{-9.5}$ and $10^{-7.0}$ to the nearest
0.5\,dex, for total hydrogen and helium mass-fractions, respectively. We note
that owing to the presence of two convection zones and the complex nature of
convective overshoot, the total masses of hydrogen and helium may not decrease
monotonically as the assumed fraction of helium at the surface decreases.
However, it is clear that if there is no helium at the surface (or in the
star), the total amount of hydrogen must be similar or smaller than our upper
limit.

\subsection{Kinematic analysis}
\DAQ\ has a fairly minor tangential speed of $29.87\pm0.09$\,\kms, however its
line-of-sight motion was found to be much larger in magnitude. To precisely
measure the total redshift of the white dwarf, we were unable to use the
discovery spectrum (covering 3,700--7,500\,\AA), as the arc calibration
spectrum was taken at a different sky position, and so the wavelength solution
is affected by some flexure of the telescope. Instead we used the February 2019
$i$-band spectrum which covers 7,300--9,100\,\AA, and contains a strong, narrow
\Ion{C}{i} line at 8,335\,\AA. The coadded $i$-band spectrum consists of two
sub-spectra where a single arc was taken between them. The RMS-scatter of the
fitted wavelength-solution is 0.024\,\AA\ corresponding to a 0.9\,\kms\
systematic uncertainty at 8,335\,\AA.

To measure the total-velocity shift of the spectrum, we first convolved our
best fitting model to a resolution of 1.43\,\AA, as measured from the
sky-emission spectrum. We then minimized the $\chi^2$-between this model and
the spectrum as a function of velocity-shift, finding $+2.7\pm5.1$\,\kms\
(including the $0.9$\,\kms\ calibration uncertainty). Inspection of the
normalised-residuals as function of velocity-shift, indicated that this level
of uncertainty is reasonable.

Considering the adopted mass of $1.140\pm0.008$\,\Msun\, radius of $6.22\pm0.08
\times 10^{-3}$\,\Rsun, and their covariance, the gravitational redshift of
\DAQ\ was found to be $116.7\pm2.3$\,\kms. Therefore the line-of-sight motion
is inferred to be $-114.0\pm5.6$\,\kms, i.e. the line-of-sight motion
essentially balances the gravitational redshift. This gives the
three-dimensional speed of \DAQ\ relative to the Sun as $117.8\pm5.4$\,\kms,
and $128.5\pm5.3$\,\kms\ relative to the local standard of rest.

To assess the statistical significance of the speed of \DAQ, we chose to
compare against the empirical velocity distribution for local white dwarfs.
While most known white dwarfs (since the release of \textit{Gaia} DR2) have
precisely measured proper-motions, few have well measured radial velocities
corrected for gravitational redshift. We therefore had to work with the
two-dimensional distribution in order to parametrize the three-dimensional
distribution. We selected our local sample as a subset of the \textit{Gaia}
catalogue of white dwarfs\cite{gentilefusilloetal19-1}, relying on stars with
parallaxes $>5$\,mas, relative parallax uncertainties $<5$\,\%, and absolute
\textit{Gaia} $G$ magnitudes with 0.5\,mag either side of \DAQ. We also
required white dwarf probabilities\cite{gentilefusilloetal19-1} of
$P_\mathrm{wd} > 0.95$. Finally we excluded objects with $v_\perp > 200$\,\kms,
as these are most likely Galactic halo stars and represent a different
distribution. This resulted in a sample size of 9,367 objects.

Assuming velocity components are drawn from Gaussian distributions with a
shared variance, then the transverse velocity $v_\perp$ is Rayleigh
distributed, and the three-dimensional speed is described by a
Maxwell-Boltzmann distribution. We fitted the dispersion, $\sigma$, to our
sample of $v_\perp$ by maximising the likelihood of the Rayleigh distribution
finding $\sigma = 30.9\pm0.2$\,\kms, however, comparing the histogram of the
data with the fitted distribution demonstrated clear disagreement particularly
in the high $v_\perp$ tail.

Instead we tried a mixture model of two Rayleigh distributions, i.e.
\begin{equation}
    P(v_\perp|\sigma_{1,2},f_1) = 
    f_1\,\frac{v_\perp}{\sigma_1^2}\,\mathrm{e}^{-v_\perp^2/2\sigma_1} +
    f_2\,\frac{v_\perp}{\sigma_2^2}\,\mathrm{e}^{-v_\perp^2/2\sigma_2},
    \label{eq:double_rayleigh}
\end{equation}
where $\sigma_{1,2}$ are the Gaussian dispersions of the two velocity
distributions ($\sigma_2 > \sigma_1$), and $f_{1,2}$ are the fractional
weightings where $f_2 = 1-f_1$. This essentially models the kinematic
distribution as containing two sub-populations (e.g. thin/thick disc or
merger/non-merger). However since the objective is simply to empirically
determine the shape of the 3D distribution, the specific interpretation is
unimportant. Again we maximised the likelihood of the parameters finding
$\sigma_1 = 24.0\pm0.5$\,\kms, $\sigma_2 = 44.0\pm1.2$\,\kms, and $f_1 =
0.72\pm0.03$, this time finding good agreement with the observed distribution
(Fig.~E\ref{fig:vdist}). The distribution on the 3D speed, $v_\mathrm{tot}$, is
correspondingly given by
\begin{equation}
    P(v_\mathrm{tot}|\sigma_{1,2},f_1) = 
    f_1\,\frac{v_\mathrm{tot}^2}{\sigma_1^3}\,\mathrm{e}^{-v_\mathrm{tot}^2/2\sigma_1} +
    f_2\,\frac{v_\mathrm{tot}^2}{\sigma_2^3}\,\mathrm{e}^{-v_\mathrm{tot}^2/2\sigma_2},
\end{equation}
where the values of $\sigma_{1,2}$, and $f_1$, are shared with those measured
for equation~(\ref{eq:double_rayleigh}). The distribution was then integrated
to infer that $128.5$\,\kms, sits at the 99th percentile
(Fig.~E\ref{fig:vdist}).

\begin{sfigure}
    \centering
    \includegraphics[width=\columnwidth]{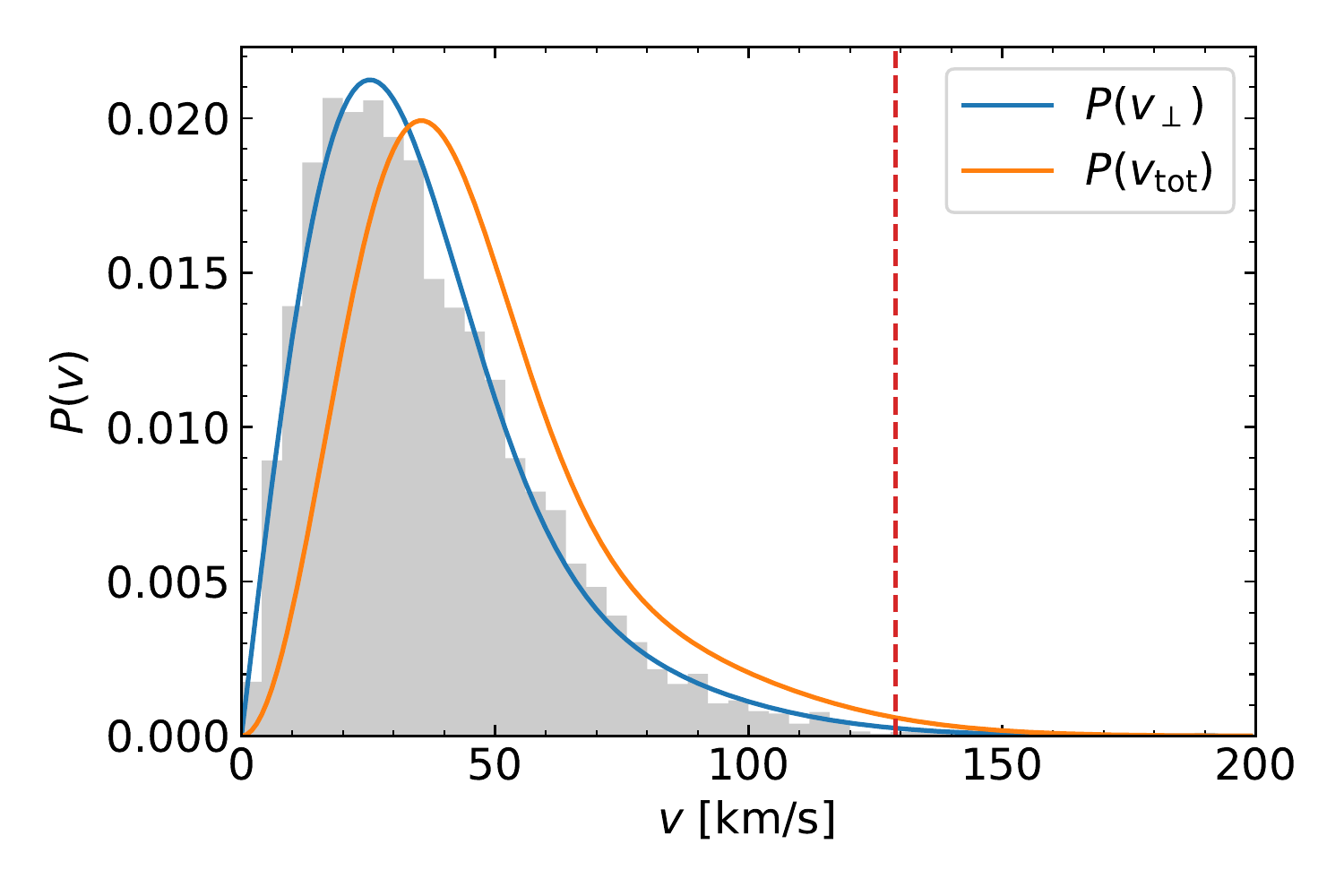}
    \caption{
    \label{fig:vdist}
    Our maximum likelihood fit (blue) to the $v_\perp$ distribution of white
    dwarfs with similar $G_\mathrm{abs}$ (grey) to \DAQ. The LSR 3D velocity of
    \DAQ\ (red dashed) is located beyond the 99th percentile of the
    corresponding 3D distribution (orange).
    }
\end{sfigure}

\subsection{Time-series photometry}

The time-series photometry shown in Fig.~\ref{fig:lcurve} were acquired using
ULTRASPEC, a high cadence photometer, mounted on the 2.4\,m Thai National
Telescope\cite{dhillonetal14-1}. Observations were made using a KG5 filter with
$2\times2$ binning. A summary of observations is given in
Table~E\ref{tab:photobslog}. The observations were bias and flat-field
corrected, and aperture photometry extracted using the \textsc{tsreduce}
pipeline\cite{choteetal14-1}. Several nearby stars were used as comparisons to
create differential light curves relative to the mean flux level. Amplitude
spectra were determined using a discrete Fourier transform of the data.

The frequencies for both runs were calculated from the centroids of the
power-spectra (as opposed to the amplitude-spectra shown in
Fig.~\ref{fig:lcurve}) in a $500$\,\uHz\ window around the main peak. I.e.
\begin{equation}
    f_c = \sum_i P_i f_i / \sum_i P_i,
\end{equation}
where $P_i$ is the power at frequency $f_i$. Similarly the centroid variances,
and hence their uncertainties were calculated as
\begin{equation}
    V = \sum_i P_i (f_i-f_c)^2 / \sum_i P_i.
\end{equation}
For the January and February observing runs, this yielded frequency centroids
of $1186.4\pm3.6$\,\uHz\ and $1202.0\pm3.2$\,\uHz\ respectively. Thus
$15.6\pm4.8$\,\uHz\ was inferred as the frequency shift between runs.

\begin{table}
    \scriptsize
	\centering
	\caption{Observing log for \DAQ\ spectroscopy.}
	\label{tab:obslog}
	\begin{tabular}{lccc}
        \hline
        Obs. date & Cen. wave. (blue/red, \AA) & $t_\mathrm{exp.}$ (s) & $N_\mathrm{exp.}$ \\
        \hline
        2018-10-15 & 4,540/6,650 & 600 & 1 \\
        2019-02-09 & 3,930/8,200 & 900 & 2 \\
        2019-09-07 & 4,540/8,200 & 900 & 4 \\
		\hline
	\end{tabular}
\end{table}
    
\begin{table}
	\centering
	\caption{Results from our spectro-photometric fit.
	Error-ranges represent $1\sigma$ uncertainties.}
	\label{tab:spectrophotofit1}
	\begin{tabular}{lc}
        \hline
        Parameter                     &                   \\
        \hline                        
        \Teff\ [K]                    & $13{,}370\pm330$ \\
        $R$ [$\times10^{-3}$\,\Rsun]  & $6.22\pm0.08   $  \\
        \logZ{He} [dex]               & $<-0.3         $  \\
        \logZ{C}  [dex]               & $-0.83\pm0.04  $  \\
        \logZ{O}  [dex]               & $<-4.5         $  \\
        \hline   
        $M_\mathrm{CO}$  [\Msun]      & $1.145\pm0.007 $  \\
        $M_\mathrm{ONe}$ [\Msun]      & $1.136\pm0.007 $  \\
        $\log g_\mathrm{CO}$  [dex]   & $8.910\pm0.014 $  \\
        $\log g_\mathrm{ONe}$ [dex]   & $8.906\pm0.014 $  \\
        $\log(L/L_\odot)$             & $-2.95\pm0.04  $  \\
		\hline
	\end{tabular}
\end{table}

\begin{table}
    \scriptsize
	\centering
	\caption{Astrometry and photometry for \DAQ.
        All astrometric data is from Gaia DR2, and thus at the J2015.5 epoch.
        Photometry is in units of magnitudes. Gaia magnitudes have been
        calculated in the AB system and include uncertainty in the Gaia
        zeropoints.
        Error-ranges represent $1\sigma$ uncertainties.}
	\label{tab:phot_ast_table}
	\begin{tabular}{lc}
        \hline
        Parameter & Value \\
        \hline
        Ra (J2015.5)  & 05:51:34.612 \\
        Dec (J2015.5) & +41:35:31.09 \\
        Gaia source ID & 192275966334956672 \\
        $\varpi$ [mas] & $21.5306\pm0.0690$ \\
        $\mu_\mathrm{Ra}$  [mas\,yr$^{-1}$] & $+114.209\pm0.117$ \\
        $\mu_\mathrm{Dec}$ [mas\,yr$^{-1}$] & $+73.207\pm0.100$ \\
        \hline
        Gaia $G$   & $16.4701\pm0.0023$ \\
        Gaia $B_p$ & $16.3503\pm0.0084$ \\
        Gaia $R_p$ & $16.7441\pm0.0045$ \\
        \hline
        Galex FUV      & $18.940\pm0.075$ \\
        Galex NUV      & $16.596\pm0.018$ \\
        Swift $U$      & $16.4030\pm0.0270$ \\
        Swift $UVW1$   & $16.3603\pm0.0276$ \\
        Swift $UVW2$   & $16.6224\pm0.0289$ \\
        Pan-STARRS $g$ & $16.3134\pm0.0045$ \\
        Pan-STARRS $r$ & $16.4689\pm0.0021$ \\
        Pan-STARRS $i$ & $16.6781\pm0.0063$ \\
        Pan-STARRS $z$ & $16.8980\pm0.0037$ \\
        Pan-STARRS $y$ & $17.0738\pm0.0098$ \\
		\hline
	\end{tabular}
\end{table}

\begin{table}
    \scriptsize
	\centering
	\caption{Observing log for TNT lightcurves of \DAQ.}
	\label{tab:photobslog}
	\begin{tabular}{lcc}
        \hline
        Obs. date & $t_\mathrm{exp.}$ (s) & $N_\mathrm{exp.}$ \\
        \hline
        2019-01-24 & 10   & 1,046 \\
        2019-01-25 & 5/10 & 902  \\
        2019-02-03 & 5    & 2,008 \\
        2019-02-04 & 5    & 1,985 \\
		\hline
	\end{tabular}
\end{table}
\end{methods}
\newpage




\noindent\textbf{Data Availability Statement}\quad The spectra of \DAQ, the
best fitting model spectrum, and lightcurves are provided as supplementary
data.
\newline
\noindent\textbf{Code Availability Statement}\quad The Koester model atmosphere
and envelope codes, as well as the \textsc{lpcode}/\textsc{lp-pul}
evolutionary/pulsation codes are not made available. However their associated
references in the main text can be consulted for further details.

\section*{References}
\bibliographystyle{naturemag}
\bibliography{aanat,aabib,tremblaybib}

\begin{thebibliography}{10}
\expandafter\ifx\csname url\endcsname\relax
  \def\url#1{\texttt{#1}}\fi
\expandafter\ifx\csname urlprefix\endcsname\relax\def\urlprefix{URL }\fi
\providecommand{\bibinfo}[2]{#2}
\providecommand{\eprint}[2][]{\url{#2}}

\bibitem{pacz70}
\bibinfo{author}{{Paczy{\'n}ski}, B.}
\newblock \bibinfo{title}{{Evolution of Single Stars. I. Stellar Evolution from
  Main Sequence to White Dwarf or Carbon Ignition}}.
\newblock \emph{\bibinfo{journal}{Acta Astron.}} \textbf{\bibinfo{volume}{20}},
  \bibinfo{pages}{47} (\bibinfo{year}{1970}).

\bibitem{camisassa19}
\bibinfo{author}{{Camisassa}, M.~E.} \emph{et~al.}
\newblock \bibinfo{title}{{The evolution of ultra-massive white dwarfs}}.
\newblock \emph{\bibinfo{journal}{Astron. Astrophys.}}
  \textbf{\bibinfo{volume}{625}}, \bibinfo{pages}{A87} (\bibinfo{year}{2019}).
\newblock \eprint{1807.03894}.

\bibitem{hollandsetal18-2}
\bibinfo{author}{{Hollands}, M.~A.}, \bibinfo{author}{{Tremblay}, P.~E.},
  \bibinfo{author}{{G{\"a}nsicke}, B.~T.}, \bibinfo{author}{{Gentile-Fusillo},
  N.~P.} \& \bibinfo{author}{{Toonen}, S.}
\newblock \bibinfo{title}{{The Gaia 20 pc white dwarf sample}}.
\newblock \emph{\bibinfo{journal}{Mon. Not. R. Astron. Soc.}}
  \textbf{\bibinfo{volume}{480}}, \bibinfo{pages}{3942--3961}
  (\bibinfo{year}{2018}).
\newblock \eprint{1805.12590}.

\bibitem{kepler19}
\bibinfo{author}{{Kepler}, S.~O.} \emph{et~al.}
\newblock \bibinfo{title}{{White dwarf and subdwarf stars in the Sloan Digital
  Sky Survey Data Release 14}}.
\newblock \emph{\bibinfo{journal}{Mon. Not. R. Astron. Soc.}}
  \textbf{\bibinfo{volume}{486}}, \bibinfo{pages}{2169--2183}
  (\bibinfo{year}{2019}).
\newblock \eprint{1904.01626}.

\bibitem{koester82}
\bibinfo{author}{{Koester}, D.}, \bibinfo{author}{{Weidemann}, V.} \&
  \bibinfo{author}{{Zeidler}, E.~M.}
\newblock \bibinfo{title}{{Atmospheric parameters and carbon abundance of white
  dwarfs of spectral types C2 and DC.}}
\newblock \emph{\bibinfo{journal}{Astron. Astrophys.}}
  \textbf{\bibinfo{volume}{116}}, \bibinfo{pages}{147--157}
  (\bibinfo{year}{1982}).

\bibitem{pelletier86}
\bibinfo{author}{{Pelletier}, C.}, \bibinfo{author}{{Fontaine}, G.},
  \bibinfo{author}{{Wesemael}, F.}, \bibinfo{author}{{Michaud}, G.} \&
  \bibinfo{author}{{Wegner}, G.}
\newblock \bibinfo{title}{{Carbon Pollution in Helium-rich White Dwarf
  Atmospheres: Time-dependent Calculations of the Dredge-up Process}}.
\newblock \emph{\bibinfo{journal}{Astrophys. J.}}
  \textbf{\bibinfo{volume}{307}}, \bibinfo{pages}{242} (\bibinfo{year}{1986}).

\bibitem{coutu19}
\bibinfo{author}{{Coutu}, S.} \emph{et~al.}
\newblock \bibinfo{title}{{Analysis of Helium-rich White Dwarfs Polluted by
  Heavy Elements in the Gaia Era}}.
\newblock \emph{\bibinfo{journal}{Astrophys. J.}}
  \textbf{\bibinfo{volume}{885}}, \bibinfo{pages}{74} (\bibinfo{year}{2019}).

\bibitem{koester19}
\bibinfo{author}{{Koester}, D.} \& \bibinfo{author}{{Kepler}, S.~O.}
\newblock \bibinfo{title}{{Carbon-rich (DQ) white dwarfs in the Sloan Digital
  Sky Survey}}.
\newblock \emph{\bibinfo{journal}{Astron. Astrophys.}}
  \textbf{\bibinfo{volume}{628}}, \bibinfo{pages}{A102} (\bibinfo{year}{2019}).
\newblock \eprint{1905.11174}.

\bibitem{rolland18}
\bibinfo{author}{{Rolland}, B.}, \bibinfo{author}{{Bergeron}, P.} \&
  \bibinfo{author}{{Fontaine}, G.}
\newblock \bibinfo{title}{{On the Spectral Evolution of Helium-atmosphere White
  Dwarfs Showing Traces of Hydrogen}}.
\newblock \emph{\bibinfo{journal}{Astrophys. J.}}
  \textbf{\bibinfo{volume}{857}}, \bibinfo{pages}{56} (\bibinfo{year}{2018}).
\newblock \eprint{1803.05965}.

\bibitem{iben1983}
\bibinfo{author}{{Iben}, J., I.} \& \bibinfo{author}{{Renzini}, A.}
\newblock \bibinfo{title}{{Asymptotic giant branch evolution and beyond.}}
\newblock \emph{\bibinfo{journal}{Annu. Rev. Astron. Astrophys.}}
  \textbf{\bibinfo{volume}{21}}, \bibinfo{pages}{271--342}
  (\bibinfo{year}{1983}).

\bibitem{toonen12}
\bibinfo{author}{{Toonen}, S.}, \bibinfo{author}{{Nelemans}, G.} \&
  \bibinfo{author}{{Portegies Zwart}, S.}
\newblock \bibinfo{title}{{Supernova Type Ia progenitors from merging double
  white dwarfs. Using a new population synthesis model}}.
\newblock \emph{\bibinfo{journal}{Astron. Astrophys.}}
  \textbf{\bibinfo{volume}{546}}, \bibinfo{pages}{A70} (\bibinfo{year}{2012}).
\newblock \eprint{1208.6446}.

\bibitem{shen12}
\bibinfo{author}{{Shen}, K.~J.}, \bibinfo{author}{{Bildsten}, L.},
  \bibinfo{author}{{Kasen}, D.} \& \bibinfo{author}{{Quataert}, E.}
\newblock \bibinfo{title}{{The Long-term Evolution of Double White Dwarf
  Mergers}}.
\newblock \emph{\bibinfo{journal}{Astrophys. J.}}
  \textbf{\bibinfo{volume}{748}}, \bibinfo{pages}{35} (\bibinfo{year}{2012}).
\newblock \eprint{1108.4036}.

\bibitem{cheng19}
\bibinfo{author}{{Cheng}, S.}, \bibinfo{author}{{Cummings}, J.~D.} \&
  \bibinfo{author}{{M{\'e}nard}, B.}
\newblock \bibinfo{title}{{A Cooling Anomaly of High-mass White Dwarfs}}.
\newblock \emph{\bibinfo{journal}{Astrophys. J.}}
  \textbf{\bibinfo{volume}{886}}, \bibinfo{pages}{100} (\bibinfo{year}{2019}).

\bibitem{Gvaramadze19}
\bibinfo{author}{{Gvaramadze}, V.~V.} \emph{et~al.}
\newblock \bibinfo{title}{{A massive white-dwarf merger product before final
  collapse}}.
\newblock \emph{\bibinfo{journal}{Nature}} \textbf{\bibinfo{volume}{569}},
  \bibinfo{pages}{684--687} (\bibinfo{year}{2019}).
\newblock \eprint{1904.00012}.

\bibitem{gentilefusilloetal19-1}
\bibinfo{author}{{Gentile Fusillo}, N.~P.} \emph{et~al.}
\newblock \bibinfo{title}{{A Gaia Data Release 2 catalogue of white dwarfs and
  a comparison with SDSS}}.
\newblock \emph{\bibinfo{journal}{Mon. Not. R. Astron. Soc.}}
  \textbf{\bibinfo{volume}{482}}, \bibinfo{pages}{4570--4591}
  (\bibinfo{year}{2019}).
\newblock \eprint{1807.03315}.

\bibitem{gaia1}
\bibinfo{author}{{Gaia Collaboration}} \emph{et~al.}
\newblock \bibinfo{title}{{Gaia Data Release 2. Summary of the contents and
  survey properties}}.
\newblock \emph{\bibinfo{journal}{Astron. Astrophys.}}
  \textbf{\bibinfo{volume}{616}}, \bibinfo{pages}{A1} (\bibinfo{year}{2018}).
\newblock \eprint{1804.09365}.

\bibitem{M-R}
\bibinfo{author}{{Chandrasekhar}, S.}
\newblock \bibinfo{title}{{The highly collapsed configurations of a stellar
  mass (Second paper)}}.
\newblock \emph{\bibinfo{journal}{Mon. Not. R. Astron. Soc.}}
  \textbf{\bibinfo{volume}{95}}, \bibinfo{pages}{207--225}
  (\bibinfo{year}{1935}).

\bibitem{sionetal83-1}
\bibinfo{author}{{Sion}, E.~M.} \emph{et~al.}
\newblock \bibinfo{title}{{A proposed new white dwarf spectral classification
  system}}.
\newblock \emph{\bibinfo{journal}{Astrophys. J.}}
  \textbf{\bibinfo{volume}{269}}, \bibinfo{pages}{253--257}
  (\bibinfo{year}{1983}).

\bibitem{althausetal12-1}
\bibinfo{author}{{Althaus}, L.~G.}, \bibinfo{author}{{Garc{\'\i}a-Berro}, E.},
  \bibinfo{author}{{Isern}, J.}, \bibinfo{author}{{C{\'o}rsico}, A.~H.} \&
  \bibinfo{author}{{Miller Bertolami}, M.~M.}
\newblock \bibinfo{title}{{New phase diagrams for dense carbon-oxygen mixtures
  and white dwarf evolution}}.
\newblock \emph{\bibinfo{journal}{Astron. Astrophys.}}
  \textbf{\bibinfo{volume}{537}}, \bibinfo{pages}{A33} (\bibinfo{year}{2012}).
\newblock \eprint{1110.5665}.

\bibitem{koester09-1}
\bibinfo{author}{{Koester}, D.}
\newblock \bibinfo{title}{Accretion and diffusion in white dwarfs. new
  diffusion timescales and applications to gd 362 and g 29-38}.
\newblock \emph{\bibinfo{journal}{Astron. Astrophys.}}
  \textbf{\bibinfo{volume}{498}}, \bibinfo{pages}{517--525}
  (\bibinfo{year}{2009}).
\newblock \eprint{0903.1499}.

\bibitem{cunningham2019}
\bibinfo{author}{{Cunningham}, T.}, \bibinfo{author}{{Tremblay}, P.-E.},
  \bibinfo{author}{{Freytag}, B.}, \bibinfo{author}{{Ludwig}, H.-G.} \&
  \bibinfo{author}{{Koester}, D.}
\newblock \bibinfo{title}{{Convective overshoot and macroscopic diffusion in
  pure-hydrogen-atmosphere white dwarfs}}.
\newblock \emph{\bibinfo{journal}{Mon. Not. R. Astron. Soc.}}
  \textbf{\bibinfo{volume}{488}}, \bibinfo{pages}{2503--2522}
  (\bibinfo{year}{2019}).
\newblock \eprint{1906.11252}.

\bibitem{cummingsetal18-1}
\bibinfo{author}{{Cummings}, J.~D.}, \bibinfo{author}{{Kalirai}, J.~S.},
  \bibinfo{author}{{Tremblay}, P.~E.}, \bibinfo{author}{{Ramirez-Ruiz}, E.} \&
  \bibinfo{author}{{Choi}, J.}
\newblock \bibinfo{title}{{The White Dwarf Initial-Final Mass Relation for
  Progenitor Stars from 0.85 to 7.5 M$_{\odot}$}}.
\newblock \emph{\bibinfo{journal}{Astrophys. J.}}
  \textbf{\bibinfo{volume}{866}}, \bibinfo{pages}{21} (\bibinfo{year}{2018}).
\newblock \eprint{1809.01673}.

\bibitem{tremblayetal15-2}
\bibinfo{author}{{Tremblay}, P.~E.} \emph{et~al.}
\newblock \bibinfo{title}{{3D Model Atmospheres for Extremely Low-mass White
  Dwarfs}}.
\newblock \emph{\bibinfo{journal}{Astrophys. J.}}
  \textbf{\bibinfo{volume}{809}}, \bibinfo{pages}{148} (\bibinfo{year}{2015}).
\newblock \eprint{1507.01927}.

\bibitem{curdetal17-1}
\bibinfo{author}{{Curd}, B.} \emph{et~al.}
\newblock \bibinfo{title}{{Four new massive pulsating white dwarfs including an
  ultramassive DAV}}.
\newblock \emph{\bibinfo{journal}{Mon. Not. R. Astron. Soc.}}
  \textbf{\bibinfo{volume}{468}}, \bibinfo{pages}{239--249}
  (\bibinfo{year}{2017}).
\newblock \eprint{1702.03343}.

\bibitem{dufouretal07-1}
\bibinfo{author}{{Dufour}, P.}, \bibinfo{author}{{Liebert}, J.},
  \bibinfo{author}{{Fontaine}, G.} \& \bibinfo{author}{{Behara}, N.}
\newblock \bibinfo{title}{White dwarf stars with carbon atmospheres}.
\newblock \emph{\bibinfo{journal}{Nature}} \textbf{\bibinfo{volume}{450}},
  \bibinfo{pages}{522--524} (\bibinfo{year}{2007}).
\newblock \eprint{0711.3227}.

\bibitem{ferrarioetal97-1}
\bibinfo{author}{{Ferrario}, L.}, \bibinfo{author}{{Vennes}, S.},
  \bibinfo{author}{{Wickramasinghe}, D.~T.}, \bibinfo{author}{{Bailey}, J.~A.}
  \& \bibinfo{author}{{Christian}, D.~J.}
\newblock \bibinfo{title}{Euve j0317-855 a rapidly rotating, high-field
  magnetic white dwarf}.
\newblock \emph{\bibinfo{journal}{Mon. Not. R. Astron. Soc.}}
  \textbf{\bibinfo{volume}{292}}, \bibinfo{pages}{205--217}
  (\bibinfo{year}{1997}).

\bibitem{corsico06}
\bibinfo{author}{{C{\'o}rsico}, A.~H.} \& \bibinfo{author}{{Althaus}, L.~G.}
\newblock \bibinfo{title}{{Asteroseismic inferences on GW Virginis variable
  stars in the frame of new PG 1159 evolutionary models}}.
\newblock \emph{\bibinfo{journal}{Astron. Astrophys.}}
  \textbf{\bibinfo{volume}{454}}, \bibinfo{pages}{863--881}
  (\bibinfo{year}{2006}).
\newblock \eprint{astro-ph/0603736}.

\bibitem{degeronimoetal19-1}
\bibinfo{author}{{De Ger{\'o}nimo}, F.~C.}, \bibinfo{author}{{C{\'o}rsico},
  A.~H.}, \bibinfo{author}{{Althaus}, L.~G.}, \bibinfo{author}{{Wachlin},
  F.~C.} \& \bibinfo{author}{{Camisassa}, M.~E.}
\newblock \bibinfo{title}{{Pulsation properties of ultra-massive DA white dwarf
  stars with ONe cores}}.
\newblock \emph{\bibinfo{journal}{Astron. Astrophys.}}
  \textbf{\bibinfo{volume}{621}}, \bibinfo{pages}{A100} (\bibinfo{year}{2019}).
\newblock \eprint{1807.03810}.

\bibitem{corsicoetal19-1}
\bibinfo{author}{{C{\'o}rsico}, A.~H.}, \bibinfo{author}{{Althaus}, L.~G.},
  \bibinfo{author}{{Miller Bertolami}, M.~M.} \& \bibinfo{author}{{Kepler},
  S.~O.}
\newblock \bibinfo{title}{{Pulsating white dwarfs: new insights}}.
\newblock \emph{\bibinfo{journal}{Astron. Astrophys. Rev.}}
  \textbf{\bibinfo{volume}{27}}, \bibinfo{pages}{7} (\bibinfo{year}{2019}).
\newblock \eprint{1907.00115}.

\bibitem{Tanikawa15}
\bibinfo{author}{{Tanikawa}, A.} \emph{et~al.}
\newblock \bibinfo{title}{{Hydrodynamical Evolution of Merging Carbon-Oxygen
  White Dwarfs: Their Pre-supernova Structure and Observational Counterparts}}.
\newblock \emph{\bibinfo{journal}{Astrophys. J.}}
  \textbf{\bibinfo{volume}{807}}, \bibinfo{pages}{40} (\bibinfo{year}{2015}).
\newblock \eprint{1504.06035}.

\bibitem{currieetal14-1}
\bibinfo{author}{{Currie}, M.~J.} \emph{et~al.}
\newblock \emph{\bibinfo{title}{{Starlink Software in 2013}}}, vol.
  \bibinfo{volume}{485} of \emph{\bibinfo{series}{Astron. Soc. Pac. Conf.}},
  \bibinfo{pages}{391} (\bibinfo{year}{2014}).

\bibitem{marsh19-1}
\bibinfo{author}{{Marsh}, T.~R.}
\newblock \bibinfo{title}{{molly: 1D astronomical spectra analyzer}}.
\newblock \emph{\bibinfo{journal}{Astrophys. Source Code Libr.}}
  \bibinfo{pages}{1907.012} (\bibinfo{year}{2019}).

\bibitem{koester10-1}
\bibinfo{author}{{Koester}, D.}
\newblock \bibinfo{title}{White dwarf spectra and atmosphere models .}
\newblock \emph{\bibinfo{journal}{Memorie della Societa Astronomica Italiana,}}
  \textbf{\bibinfo{volume}{81}}, \bibinfo{pages}{921--931}
  (\bibinfo{year}{2010}).

\bibitem{gaia2}
\bibinfo{author}{{Gaia Collaboration}} \emph{et~al.}
\newblock \bibinfo{title}{{Gaia Data Release 2. Observational
  Hertzsprung-Russell diagrams}}.
\newblock \emph{\bibinfo{journal}{Astron. Astrophys.}}
  \textbf{\bibinfo{volume}{616}}, \bibinfo{pages}{A10} (\bibinfo{year}{2018}).
\newblock \eprint{1804.09378}.

\bibitem{Pan-STARSS1}
\bibinfo{author}{{Chambers}, K.~C.} \emph{et~al.}
\newblock \bibinfo{title}{{The Pan-STARRS1 Surveys}}.
\newblock \emph{\bibinfo{journal}{arXiv e-prints}}
  \bibinfo{pages}{arXiv:1612.05560} (\bibinfo{year}{2016}).
\newblock \eprint{1612.05560}.

\bibitem{galex}
\bibinfo{author}{{Morrissey}, P.} \emph{et~al.}
\newblock \bibinfo{title}{{The Calibration and Data Products of GALEX}}.
\newblock \emph{\bibinfo{journal}{Astrophys. J. Suppl.}}
  \textbf{\bibinfo{volume}{173}}, \bibinfo{pages}{682--697}
  (\bibinfo{year}{2007}).

\bibitem{walletal19-1}
\bibinfo{author}{{Wall}, R.~E.} \emph{et~al.}
\newblock \bibinfo{title}{{GALEX absolute calibration and extinction
  coefficients based on white dwarfs}}.
\newblock \emph{\bibinfo{journal}{Mon. Not. R. Astron. Soc.}}
  \textbf{\bibinfo{volume}{489}}, \bibinfo{pages}{5046--5052}
  (\bibinfo{year}{2019}).
\newblock \eprint{1909.02617}.

\bibitem{tremblayetal13-1}
\bibinfo{author}{{Tremblay}, P.-E.}, \bibinfo{author}{{Ludwig}, H.-G.},
  \bibinfo{author}{{Steffen}, M.} \& \bibinfo{author}{{Freytag}, B.}
\newblock \bibinfo{title}{{Spectroscopic analysis of DA white dwarfs with 3D
  model atmospheres}}.
\newblock \emph{\bibinfo{journal}{Astron. Astrophys.}}
  \textbf{\bibinfo{volume}{559}}, \bibinfo{pages}{A104} (\bibinfo{year}{2013}).
\newblock \eprint{1309.0886}.

\bibitem{fontaineetal01-1}
\bibinfo{author}{{Fontaine}, G.}, \bibinfo{author}{{Brassard}, P.} \&
  \bibinfo{author}{{Bergeron}, P.}
\newblock \bibinfo{title}{{The Potential of White Dwarf Cosmochronology}}.
\newblock \emph{\bibinfo{journal}{Publ. Astron. Soc. Pac.}}
  \textbf{\bibinfo{volume}{113}}, \bibinfo{pages}{409--435}
  (\bibinfo{year}{2001}).

\bibitem{siess07-1}
\bibinfo{author}{{Siess}, L.}
\newblock \bibinfo{title}{{Evolution of massive AGB stars. II. model properties
  at non-solar metallicity and the fate of Super-AGB stars}}.
\newblock \emph{\bibinfo{journal}{Astron. Astrophys.}}
  \textbf{\bibinfo{volume}{476}}, \bibinfo{pages}{893--909}
  (\bibinfo{year}{2007}).

\bibitem{siess10-1}
\bibinfo{author}{{Siess}, L.}
\newblock \bibinfo{title}{{Evolution of massive AGB stars. III. the thermally
  pulsing super-AGB phase}}.
\newblock \emph{\bibinfo{journal}{Astron. Astrophys.}}
  \textbf{\bibinfo{volume}{512}}, \bibinfo{pages}{A10} (\bibinfo{year}{2010}).

\bibitem{paquette1}
\bibinfo{author}{{Paquette}, C.}, \bibinfo{author}{{Pelletier}, C.},
  \bibinfo{author}{{Fontaine}, G.} \& \bibinfo{author}{{Michaud}, G.}
\newblock \bibinfo{title}{{Diffusion coefficients for stellar plasmas}}.
\newblock \emph{\bibinfo{journal}{Astrophys. J. Suppl.}}
  \textbf{\bibinfo{volume}{61}}, \bibinfo{pages}{177--195}
  (\bibinfo{year}{1986}).

\bibitem{paquette2}
\bibinfo{author}{{Paquette}, C.}, \bibinfo{author}{{Pelletier}, C.},
  \bibinfo{author}{{Fontaine}, G.} \& \bibinfo{author}{{Michaud}, G.}
\newblock \bibinfo{title}{{Diffusion in white dwarfs - New results and
  comparative study}}.
\newblock \emph{\bibinfo{journal}{Astrophys. J. Suppl.}}
  \textbf{\bibinfo{volume}{61}}, \bibinfo{pages}{197--217}
  (\bibinfo{year}{1986}).

\bibitem{iben+mcdonald85-1}
\bibinfo{author}{{Iben}, J., I.} \& \bibinfo{author}{{MacDonald}, J.}
\newblock \bibinfo{title}{{The effects of diffusion due to gravity and due to
  composition gradients on the rate of hydrogen burning in a cooling degenerate
  dwarf. I - The case of a thick helium buffer layer}}.
\newblock \emph{\bibinfo{journal}{Astrophys. J.}}
  \textbf{\bibinfo{volume}{296}}, \bibinfo{pages}{540--553}
  (\bibinfo{year}{1985}).

\bibitem{overshoot1}
\bibinfo{author}{{Spiegel}, E.~A.}
\newblock \bibinfo{title}{{A Generalization of the Mixing-Length Theory of
  Turbulent Convection.}}
\newblock \emph{\bibinfo{journal}{Astrophys. J.}}
  \textbf{\bibinfo{volume}{138}}, \bibinfo{pages}{216} (\bibinfo{year}{1963}).

\bibitem{overshoot2}
\bibinfo{author}{{Zahn}, J.-P.}
\newblock \bibinfo{title}{{Convective penetration in stellar interiors}}.
\newblock \emph{\bibinfo{journal}{Astron. Astrophys.}}
  \textbf{\bibinfo{volume}{252}}, \bibinfo{pages}{179--188}
  (\bibinfo{year}{1991}).

\bibitem{tremblayetal15-3}
\bibinfo{author}{{Tremblay}, P.~E.} \emph{et~al.}
\newblock \bibinfo{title}{{Calibration of the Mixing-length Theory for
  Convective White Dwarf Envelopes}}.
\newblock \emph{\bibinfo{journal}{Astrophys. J.}}
  \textbf{\bibinfo{volume}{799}}, \bibinfo{pages}{142} (\bibinfo{year}{2015}).
\newblock \eprint{1412.1789}.

\bibitem{kupka18}
\bibinfo{author}{{Kupka}, F.}, \bibinfo{author}{{Zaussinger}, F.} \&
  \bibinfo{author}{{Montgomery}, M.~H.}
\newblock \bibinfo{title}{{Mixing and overshooting in surface convection zones
  of DA white dwarfs: first results from ANTARES}}.
\newblock \emph{\bibinfo{journal}{Mon. Not. R. Astron. Soc.}}
  \textbf{\bibinfo{volume}{474}}, \bibinfo{pages}{4660--4671}
  (\bibinfo{year}{2018}).
\newblock \eprint{1712.00641}.

\bibitem{cukanovaite2019}
\bibinfo{author}{{Cukanovaite}, E.} \emph{et~al.}
\newblock \bibinfo{title}{{Calibration of the mixing-length theory for
  structures of helium-dominated atmosphere white dwarfs}}.
\newblock \emph{\bibinfo{journal}{Mon. Not. R. Astron. Soc.}}
  \textbf{\bibinfo{volume}{490}}, \bibinfo{pages}{1010--1025}
  (\bibinfo{year}{2019}).
\newblock \eprint{1909.10532}.

\bibitem{dhillonetal14-1}
\bibinfo{author}{{Dhillon}, V.~S.} \emph{et~al.}
\newblock \bibinfo{title}{{ULTRASPEC: a high-speed imaging photometer on the
  2.4-m Thai National Telescope}}.
\newblock \emph{\bibinfo{journal}{Mon. Not. R. Astron. Soc.}}
  \textbf{\bibinfo{volume}{444}}, \bibinfo{pages}{4009--4021}
  (\bibinfo{year}{2014}).
\newblock \eprint{1408.2733}.

\bibitem{choteetal14-1}
\bibinfo{author}{{Chote}, P.} \emph{et~al.}
\newblock \bibinfo{title}{{Puoko-nui: a flexible high-speed photometric
  system}}.
\newblock \emph{\bibinfo{journal}{Mon. Not. R. Astron. Soc.}}
  \textbf{\bibinfo{volume}{440}}, \bibinfo{pages}{1490--1497}
  (\bibinfo{year}{2014}).
\newblock \eprint{1412.5724}.

\bibitem{tremblayetal19-1}
\bibinfo{author}{{Tremblay}, P.-E.} \emph{et~al.}
\newblock \bibinfo{title}{{Core crystallization and pile-up in the cooling
  sequence of evolving white dwarfs}}.
\newblock \emph{\bibinfo{journal}{Nature}} \textbf{\bibinfo{volume}{565}},
  \bibinfo{pages}{202--205} (\bibinfo{year}{2019}).

\end{thebibliography}

\begin{addendum}
 \item M.A.H. acknowledges useful discussions on the nature of \DAQ\ with Pier
 Bergeron and Amanda Karakas, and Maria-Teresa Belmonte for a useful discussion
 on the quality of experimental atomic data. The research leading to these
 results has received funding from the European Research Council under the
 European Union's Horizon 2020 research and innovation programme no. 677706
 (WD3D). A.A. acknowledges generous support from Faculty of Science, Naresuan
 University (grant no. R2562E029). V.S.D. and ULTRASPEC are funded by the STFC.
 This work presents results from the European Space Agency (ESA) space mission
 Gaia. Gaia data are being processed by the Gaia Data Processing and Analysis
 Consortium (DPAC). Funding for the DPAC is provided by national institutions,
 in particular the institutions participating in the Gaia MultiLateral
 Agreement (MLA). The Gaia mission website is https://www.cosmos.esa.int/gaia.
 The Gaia archive website is https://archives.esac.esa.int/gaia. The William
 Herschel Telescope is operated on the island of La Palma by the Isaac Newton
 Group of Telescopes in the Spanish Observatorio del Roque de los Muchachos of
 the Instituto de Astrofísica de Canarias. Based on observations made with
 ULTRASPEC at the Thai National Observatory which is operated by the National
 Astronomical Research Institute of Thailand (Public Organization).
 \item[Author Contribution]
 M.A.H., P.-E.T., and B.T.G. lead the project including the interpretation of \DAQ.
 M.E.C. calculated the interior CO/ONe core models.
 D.K. calculated the envelope models and advised M.A.H on atmospheric modelling.
 N.P.G.-F. acquired the initial LT lightcurve.
 A.A., V.S.D., and T.R.M. acquired the TNT lightcurves.
 P.C. calibrated the LT and TNT lightcurves and their amplitude spectra.
 A.H.C. calculated the pulsation properties of \DAQ\ from the CO/ONe interior models.
 M.J.H. and P.I. acquired the WHT spectroscopic data of \DAQ.
 D.S. acquired and calibrated SWIFT photometry of \DAQ.
 \item[Competing Interests] The authors declare that they have no competing
 financial interests.
 \item[Correspondence] Correspondence and requests for materials should be
 addressed to M. A. Hollands.~(email: M.Hollands.1@warwick.ac.uk).
\end{addendum}

\end{document}